\documentclass[article,10pt]{wlpeerj}
\usepackage{amssymb}
\usepackage{amsmath}

\title{Economic Analysis of Ransomware}

\author[1]{Julio Hernandez-Castro}
\author[2]{Edward Cartwright}
\author[2]{Anna Stepanova}
\affil[1]{School of Computing, Cornwallis South, University of Kent, UK}
\affil[2]{School of Economics, Keynes College, University of Kent, UK}


\begin{abstract}
We present in this work an economic analysis of ransomware, with relevant data from Cryptolocker, CryptoWall, TeslaCrypt and other major strands. We include a detailed study of the impact that different price discrimination strategies can have on the success of a ransomware family, examining uniform pricing, optimal price discrimination and bargaining strategies and analysing their advantages and limitations. In addition, we present results of a preliminary survey that can helps in estimating an optimal ransom value. We discuss at each stage whether the different schemes we analyse have been encountered already in existing malware, and the likelihood of them being implemented and becoming successful. We hope this work will help to gain some useful insights for predicting how ransomware may evolve in the future and be better prepared to counter its current and future threat.

\end{abstract}

\begin{document}

\flushbottom
\maketitle
\thispagestyle{empty}

\section*{Introduction}

The term ransomware, added only in 2012 to the Oxford English Dictionary (Jouhal, 2017) denotes\footnote{The Oxford definition is \textit{A type of malicious software designed to block access to a computer system until a sum of money is paid}} the branch of malware that, after infecting a computer, asks for a ransom. The name is too general and denotes all kinds of extortions, including malware that puts compromising or illegal materials in your computer and asks for a ransom not to report to the Police or to divulge this publicly. This type of malware has also been named policeware.\\ 

We are, however, more interested in our work on a more subtle kind of malware, originally called cryptovirus but later also referred to as crypto-ransomware or simply ransomware, where the malware typically encrypts and then deletes your original data files, and asks for a ransom to hand them back to you. \\

The original concept of cryptovirus was first presented in the academic literature by Adam Young and Prof. Moti Yung from Columbia University, around 1996 (Young and Yung, 1996). It was likely inspired by previous unsuccessful attempts to extort money out of infected computers by, between others, the AIDS malware. \\

The key development in the Young \& Yung approach was to employ public key cryptography for performing this extortion in a cryptographically sound and robust\footnote{Robust here means that this scheme is not vulnerable to key compromise by reverse-engineering, as so frequently occurred with substandard malware using symmetric key, because there is no impact for the criminal if the public key is made public.} manner. After their initial publications, the concept has been well known for a while by academics but only been used scarcely by malware developers, who had apparently stumbled upon it and fine-tuned it (Emm, 2008) mostly by trial and error\footnote{The case of the Gpcode ransomware is particularly interesting in this regard, in that we can see a criminal experimenting and making multiple mistakes and numerous failed attempts before getting it right in the end.}.

\begin{figure}[!h]
 \begin{center}
 \includegraphics[width=12cm]{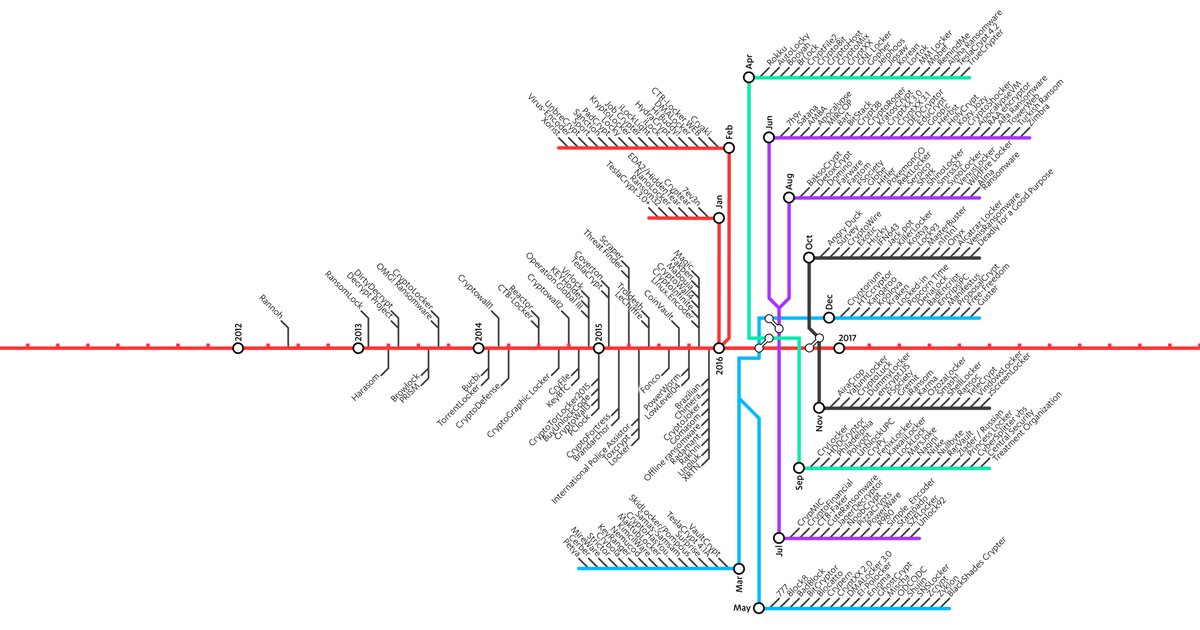} 
 \end{center}
 \caption{Ransomware Tubemap, from (F-Secure, 2017)}
 \label{fig:tubemap}
 \end{figure}

The infamous cryptolocker was one of the first, if not the first, to implement a scheme close to the Young \& Yung protocol in a technically sound way, from its conception. It was discovered in the wild in 2013 and since then there has been an explosion on the number of different ransomware families and variants (see Figure~\ref{fig:tubemap}) found, building up an industry that was recently estimated to generate 1 billion profit in 2016 by the FBI\footnote{\textit{The FBI has announced that ransomware could become a \$1 billion dollar industry, after early estimates of ransomware losses from only the first quarter of 2016 eclipsed that of 2015.} wrote Max Metzger on January 10, 2017 for SC Magazine UK.}. 

%Put image of the map ransomware explosion, find the reference to the 1 billion

The task of developing a cryptovirus from scratch is technically challenging, with many opportunities to go wrong, particularly for the cryptography illiterate. 

It is surprising how many still commit the fundamental sin of using private key cryptography, which is easily reversed as the binaries have to code the ‘secret’ key that can, thus, be easily retrieved by skilled reverse engineers. 

But even if criminals were familiar with the academic work on cryptovirus, the proper development and implementation of such a complex piece of software is still fraught with multiple opportunities to incur into fatal technical errors that would defeat the profit-making purpose of its authors. 

We have seen some recent examples where means for recovering the key or decrypting the files were quickly found and shared by the security community, leading to negligible business for the cyber criminals\footnote{A good example of this flawed ransomware is TorLocker, active in 2014 and first targeted at Japanese users, in which despite all files being encrypted with AES-256 and RSA-2048, in more than 70\% of the cases the key can be obtained and the files decrypted due to errors made during the implementation of cryptography algorithms (Sinitsyn, 2015). But there are many others, such as Cryptear, Hidden-Tear, EDA2, Stampado, NoobCrypt, Crypt0, 777, Petya, KeRanger, CryptoDefense, earlier versions of Torrentlocker and Jigsaw, and dozens more (Hahn, 2016)}
thanks to programming errors or plain ignorance, which fortunately led to a complete failure to cash-in by the wrong-doers.

\subsection*{Cryptolocker et al.}

Cryptolocker has been the first of the modern strand of cryptovirus, and one of the most successful so far in part due to its good implementation and security, which unfortunately forced victims wanting to recover their files to pay. That was the only available alternative. But even in the case of paying a ransom, victims were not completely sure of recovering their files (we estimate the recovery rate to be around 65\%)\footnote{See the results of the \textit{2016 Kent Cyber Security Survey} by Hernandez-Castro, J. and Boiten, E. at https://cyber.kent.ac.uk/Survey2016.pdf}

%cite our survey research here

That rate was high enough that victims (by press reports, etc.) knew that the criminals had a decent reputation of honoring payments with returning the key. 
Cryptolocker has mostly been distributed by email attachments and through the GameBot/Zeus botnet. Botnets are large sets of potentially millions of infected computers that can be controlled centrally by a criminal. Creating large botnets is not particularly difficult nowadays, with so many unpatched windows machines, but monetizing these infected machines has proven harder. Typical approaches consisted in selling the networking bandwidth for launching Distributed Denial of Service attacks for a small fee, or to abuse their computational power to mine cryptocoins. The new market 'opportunity' of turning them into ransomware victims is a relatively new development but one that is likely to become much more popular due to its high\footnote{In the \textit{2016 Kent Cyber Security Survey} of UK users, 26\% of the infected victims acknowledged they paid the ransom. The previous Kent Cyber Security Survey, focused exclusively in Cryptolocker pointed to an even higher conversion rate of around 35\%. Other authors claim rates that, depending on the ransomware family and the type of target can achieve anything from a 2-40\% rate.} conversion rates. 

Naturally, after the widely reported \textit{success} of Cryptolocker, many other copycats\footnote{Copycats are common in cybercrime. One quite curious approach is for criminals to try to benefit from the reputation of a popular ransomware family, and in particular to 'impersonate' it by copying their ransom notice and other characteristics. Examples of this technique are Power Worm and TeslaCrypt 2, which used CryptoWall's ransom notes (Hahn, 2016)).} have surged. We expect many more cryptoviruses to appear in the following years, as there have been reports of readily available development kits selling in the black market and the increase in popularity of Ransomware-as-a-Service (RaaS). Until the only know solution to this threats (i.e. offline backups) is widespread, criminals will continue to use this tactic to rake easy profits. Indeed, all the evidence suggests that ransomware is on the increase and will become an ever more common threat in the years to come.   

\subsection*{Bitcoin}

The crypto currency bitcoin has played a fundamental role in the success of Cryptolocker and other recent ransomware\footnote{It is important to note that, still to this day, there is a significant but decreasing number of ransomware authors that offer, in addition to bitcoin, the possibility to pay through other schemes such as UKash and PaySafeCard. Examples of this are the TorLocker and Razy 5.0 ransomware}. It allows for relatively easy money transfers and, although by no means untraceable or completely anonymous, it is considered to be secure enough to provide a good degree of anonymity. These characteristics provide cyber criminals with a very powerful tool for profiting of their crimes, and one that law enforcement is not that accustomed to.  They can use bitcoin to defeat the classical control measures already put in place to trace, follow and stop other, better-known payment methods like bank transfers. Bitcoin is currently the most popular cryptocoin, but there is a continuous stream of new proposals, some of which claim to provide full anonymity and untraceability, and will probably be preferred in the future by cyber gangs to make law enforcement work in tracking the flux of money almost impossible.  

Cryptolocker employed a number of different bitcoin addresses to request the victims to send the money to. There is speculation that it created a new one for each victim. In any case, once the victim's bitcoins were transferred to that address, they were rapidly moved to others, and laundered using bitcoin mixers.
At least 628 of these initial addresses are known.
When further investigated, it was discovered that the bitcoin transfers very frequently visited a small number of addresses. For example, from this list of 628 at least 440 visited the address 174psvzt77NgEC373xSZWm9gYXqz4sTJjn. This single address received a total of 346,102.31357807 BTC, which is a significant amount of the total number of bitcoins in circulation (approx. 12 million) at the time of its last transaction in February 2014. The average value of this amount of bitcoins at that time was in excess of \$207m. 

%The amount of transaction per day corresponding to this address can be seen below (Figure 1). 

There are other Cryptolocker accumulator addresses like 1Facb8QnikfPUoo8WVFnyai3e1Hcov9y8T, 1CbR8da9YPZqXJJKm9ze1GYf67eKAUfXwP, 1AEoiHY23fbBn8QiJ5y6oAjrhRY1Fb85uc and, between others, 1Ne5bGjDdgmbsGdNK9ocwrQiWf8K1FTuSa.

So the general procedure is as follows: From the cryptolocker addresses given to the victims for sending their payments, the ransom goes straight to accumulator addresses like those shown above. These accumulators, after receiving around 500 BTC, then sent the money to a chain of addresses that have typically only 2 transactions each, the first one receiving the bitcoin and another one sending bitcoin to two or more addresses, splitting the incoming value\footnote{This is very common in the Bitcoin network, and is generally known as a peeling chain.}. However, the bigger part of the incoming money goes to another address following the same pattern. In addition, the times of each of these two transactions are nearly identical. With this and the sheer volume of transactions, we can safely conclude that the whole process of bitcoin distribution and laundering is completely automated.

This is a common practice between modern ransomware, to immediately launder their ill-gotten gains through well-known bitcoin laundering operations (also known as mixers or tumblers), who take a fee (around 2.5\%) for their services.  

It has been repeatedly said that bitcoin friction is ransomware's only constraint\footnote{For example, by Sean Sullivan, in F-Secure's \textit{News From the Lab} blog, on 22/02/2017, that can be accessed at https://labsblog.f-secure.com/2017/02/22/bitcoin-friction-is-ransomwares-only-constraint/}
 to extend their gains even further. This may well be true, as there could be a significant number of victims willing to pay that however lack the ability to access bitcoin, particularly in short time periods as requested by some ransomware. In any case, we believe this friction will ease and virtually disappear with the popularization of bitcoin and other cryptocurrencies.

A last difficulty with bitcoin-based ransoms is the characteristic volatility of bitcoin, that sometimes has forced criminals to adjust the ransom request in BTC to keep it in line with the real cost in dollars, sterling or euros.

\subsection*{Current State of Affairs}

After Operation Tovar, led by the US Department of Justice and the FBI but involving also Europol and the UK NCA, the Australian Federal Police and other law enforcement from the Netherlands, Japan, Germany, France, etc. together with multiple security firms and some Universities, the Gameover/Zeus botnet is considered to be closed down. 

This was one of the main distribution paths for Cryptolocker and since its take-down this particular cryptovirus can be considered to be eradicated. Other attack campaigns have followed since, but of much lesser size and impact. During Operation Tovar, a victim’s database was located, containing approximately 500,000 individuals, and this allowed to set up a site online to facilitate victims to recover their files at https://www.decryptcryptolocker.com. It is important to note that this was only possible due to the recovery of this file owned by the criminals, and not to any security weakness in the implementation of the cryptovirus itself. 
This initiative has been followed by the No More Ransom Project at https://www.nomoreransom.org/.

Cryptolocker demonstrated the huge potential to extract large amounts of money through a cryptovirus. 
%We can safely assume, therefore, that 
But Cryptolocker was definitely not the end of the story. Other large scale attacks followed, and new families such as CryptoWall, TorLocker, Fusob, Cerber, TeslaCrypt, etc continued to break havoc. And as expected, criminals are refining their techniques, not only in terms of the malware component technology but also regarding the economic tools they use to extract money from victims. In this regard Cryptolocker was relatively unsophisticated compared to some modern strands. There are nowadays so many ransomware strands (see the explosion over time in Figure~\ref{fig:tubemap}) competing for our money that we argue there is something very similar to classical evolutionary pressure at play between different ransomware species to (mostly by trial and error, and replication of the fittest) survive and multiply themselves. Slightly better strategies in ransomware will provide their authors with massive rewards, and will be promptly copied by other cybercriminals. We conjecture that future attacks will thus likely evolve so that more money can be extracted by the criminals, hence converging slowly but surely towards optimal solutions from an economic standpoint. In the following we provide an economic model that shows how this can be done.

\section*{The Economics of Ransomware}

The objective of the cyber criminals is, presumably, to maximize profit from infected computers. The profit they can make largely depends on the willingness of those attacked to pay the ransom. This, in turn, will depend on various components - how much a victim values their files, the extent to which they trust the criminals to honor their word, willingness to give money to criminals, etc. For simplicity, we can subsume these components into one single figure which can be interpreted as the maximum amount a particular victim is \textit{willing to pay} to recover their files. Different people will naturally have a different willingness to pay, and so we denote by $v_i$ the willingness to pay of person $i$. For instance, a victim who values her files at \$500 and trusts the criminals would have $v_i=500$ while a victim who values her files at \$1000 but dislikes interacting with criminals or has no confidence in the return of her files may have $v_i=0$. 

The profit of the criminals can then be summarized as

\begin{center}
$\Pi=\sum_{i=1}^{N}(p_i-c) 1_i - F$
\end{center}

\noindent where $N$ is the number of people attacked, $p_i$ is the ransom asked of person $i$, $c$ is the cost of dealing with any ransom money, $1_i$ is an indicator variable that takes value $1$ if $p_i \leq v_i$ and $0$ otherwise, and $F$ is the fixed cost of operating the malware. In the following, we shall abstract away from considering $N$ and simply take it as given that the criminals will target as many people as possible. Our focus will be on the optimal ransom to charge victims. 

\subsection*{Uniform Pricing}

In order to maximize profit, criminals must determine an optimal ransom to ask each victim. This task is complicated by the fact they do not know each person’s willingness to pay $v_i$. Assume, for now, that criminals know the distribution of $v_i$’s, and so they know what proportion of people have a ‘high’ $v_i$, a ‘low’ $v_i$, etc. Assume also that they cannot distinguish individuals in any way, and so have no idea who has high $v_i$ or low $v_i$. It is then optimal to use uniform pricing, and set the ransom to the same amount for anyone attacked, $p_i=p$ for all $i$ (Pepall, Richards and Norman 2008). This appears to be the approach currently taken by most criminals and ransomware strands.

%TODO:Add here other uniform pricing ransomware!

In the case of uniform pricing, the profit of the criminals can be written as

\begin{center}
$\Pi=(p-c) Q(p) - F$
\end{center}

\noindent where $Q(p)$ is the number of people willing to pay a ransom of value $p$. Note that $Q$ can be interpreted as the demand function detailing demand `for the criminal's services' for all possible prices. 
The optimal ransom (ignoring the fact that $Q(p)$ is an integer) is then given by the formula

\begin{center}
$\frac{p-c}{p}=-\frac{1}{\eta(p)}$
\end{center}

\noindent where $\eta(p)=\frac{p}{Q(p)} \frac{dQ(p)}{dp}$ is what is called the price elasticity of demand. 

Price elasticity measures sensitivity to changes in the ransom. It is optimal for the criminals to price where demand is elastic, i.e. $|\eta(p)|\geq1$, meaning that an increase in the amount of the ransom would lead to a more than proportionate fall in the number of people paying (Pepall, Richards and Norman 2008). 

One implication of this result, possibly counter-intuitive, is that in the case of uniform pricing criminals should set the ransom at a level where a sizable proportion of victims will not be willing to pay. Indeed, if $Q(p)$ is linear, then less than 50\% of the victims should be willing to pay the ransom if it is set optimally. 

In most realistic scenarios, the criminals do not know the \textit{a priori} distribution of $v_i$’s and so do not know $Q(p)$. But they can hypothesize that, under very mild and general assumptions, the size of price elasticity $|\eta(p)|$ is a weakly decreasing function of $p$. This makes it relatively easy for the criminals to heuristically find out whether the current ransom they are using is set too high or too low. For example, suppose the current ransom is \$300 and the criminals observe that 40\% of people pay the ransom, $Q(300) = 0.4N$. If that were the case, criminals could try increasing the ransom to \$350. Suppose that at this higher price around 30\% of victims pay, $Q(350)= 0.3N$. Then, the price elasticity can be approximated by $\eta(300)=-(300/0.4N)(0.1N/50)=-1.5$. If the criminals estimate $c$ as \$10 then they can infer\\

\begin{center}
$\frac{p-c}{p}=\frac{300-10}{300} \leq \frac{1}{1.5}=-\frac{1}{\eta(p)}$\\
\end{center}

\noindent meaning that at the current ransom price, demand is too elastic. It would, in this case, be optimal to lower the ransom below \$300. 

To give a second example: Suppose that on increasing the price to \$350 around 35\% of people continue to pay the ransom, $Q(350) = 0.35N$. In this case the price elasticity can be approximated as $\eta(300)=-0.75$. If $c$ is estimated again at around \$10 then, at the current ransom, demand is too inelastic. It would, therefore, be optimal to increase the ransom above \$300. By systematically varying the ransom over time, the criminals can learn the elasticity of demand and quickly converge on the optimal price. 

We do not have any evidence of this exact strategy being used by any ransomware yet. And we reiterate that the optimal strategy is somewhat counter-intuitive in that the criminals have to price at a level where the majority of victims will not pay the ransom. Even so, we will not be surprised to find schemes employed in the near future that actively search for information about demand, as given by $Q(p)$, and set the ransom accordingly.

\subsection*{Price Discrimination}

In the previous section, we assumed criminals could not distinguish between different victim's willingness to pay. This is a reasonable assumption, supported by many examples of ransomware strands in the wild, but limits the profits criminals can make to that of a uniform pricing monopolist. They can certainly increase their profit by price discriminating. But how to do that? 

Let us begin by noting that economists distinguish three types of price discrimination (Varian 1989). First degree (or perfect) price discrimination involves each person being given an individually tailored price. This is the `gold standard' of discrimination that criminals may ultimately achieve but seems a long way off at the moment.  

More realistically criminals can use second or third degree price discrimination. Given that third degree price discrimination is simplest to apply we shall focus on that here.\footnote{With second degree price discrimination, people are offered a price menu, quantity packages. For instance, in the ransomware context there might be a basic package of $\$100$ to recover word processing files, a better package of $\$200$ that includes photos, and a complete package of $\$300$ that includes all files. The objective of offering a menu of options is to discern between different types of victims.} The objective with third degree price discrimination is to distinguish different types of victim and give a ransom based on type. Access to a person’s files and computer can give the criminals useful information about the value of the information present, the victim's worth, or both, by using the number, type and size of files, model and age of the computer, etc. Any or all of this information may correlate with a person’s willingness to pay. If so, the criminals can categorize victims into types\footnote{Smart criminals with a large base of victims can apply machine learning, regression or clustering techniques to a series of easily extracted features in a victim's system.} and apply third degree price discrimination.

To illustrate how this may work with a very simple example, suppose the criminals can distinguish people with a `large' number of files from those with a `small' number of files\footnote{This is a very realistic, almost trivial, assumption. Of course, easily replaceable system files can be subtracted from this count.}. They could then price-discriminate by setting a ransom $p_{L}$ for anyone with a large number of files, and $p_{S}$ to anyone with a small number of
files, where $p_{L}\neq p_{S}$. For each type of victim, the optimal price is determined as explained above for uniform pricing. So, the optimal ransom for those with a large number of files is

\[
\frac{p_{L}-c}{p_{L}}=-\frac{1}{\eta _{L}(p_{L})}
\]

\noindent where $\eta _{L}$ is the price elasticity of demand for people with a large number of files. Similarly, the optimal ransom for those with a small number of files can be determined from the analogous elasticity $\eta _{S}$. 

Intuitively, we might expect those with a larger number of files to have, on average, a higher willingness to pay to recover their files. This would imply that, for any given price, demand is less elastic for those with a larger number of files, $|\eta _{S}(p_{S})|>|\eta _{L}(p_{L})|$. Thus, the criminals should charge a higher price to those with a large number of files, $p_{L}>p_{S}$. 

Charging two different prices increases the criminals profit by making the ransom more personalized. But there is no reason other than simplicity to have only two different prices. The more information the criminals can extract on expected willingness to pay the more personalized should be the ransom. By systematically varying the ransom, and by taking into account the extracted information about victims, the criminals can learn over time the optimal ransom for each `type' of victim. The more personalized the pricing becomes, the closer the criminals come to attaining the maximum achievable profit, corresponding to first degree price discrimination (Varian 1989). 

Intuitively, one reason price discrimination increases profits is because it increases the number of people paying the ransom. This works because the criminals can lower the ransom for those with a lower willingness to pay, \textit{without lowering} the ransom for those with a higher willingness to pay. Under first degree price discrimination every person will receive a ransom they are willing to pay. Indeed, the criminals can extract all the surplus by setting the personalized ransom $p_{i}$ just below their victim's willingness to pay $v_{i}$. 

It is interesting that some ransomware families are starting to implement some sort of primitive price discrimination strategies. For example, the Shade strand is using a Remote Access Trojan (RAT) to spy on their victims finances (apparently a version of TeamViewer) and estimate the ransom they can request (Atanasova, 2016). This is of course a poor way to implement price discrimination, because it is time consuming, not easily scalable and opens the criminals to potentially being traced back more easily. 

There are, of course, slightly less sophisticated examples of initial attempts at price discrimination in ransomware, and one of the most notable ones is that of the Fantom ransomware (Abrams, 2016). In the latest variant of Fantom, the ransom amount is determined by the name of the process/executable file, thus allowing for multiple simultaneous attack campaigns, targeting different types of victims and requesting different ransom sums.

Another, even less subtle way of implementing price discrimination is by targeting different spam campaigns messages (and distribution lists) to specific types of businesses and/or roles within a company. This strategy can also be used to differentiate between home users and corporate users, and target ransoms accordingly. This approach is relatively common nowadays in many ransomware variants. 

Not all ransomware is spread through spam, email, or requires an 'unsafe' user intervention to infect a device. Unfortunately, another very simple and popular attack vector for ransomware is malvertising, which consists in malicious ads that can be delivered by even popular and reputable websites (such as AOL, The New York Times, BBC, MSN, etc.) These ads will (typically by using an iframe) quietly install an exploit kit and, as part of it, a ransomware. Well-known examples of ransomware families using this strategy are the Cerber 3.0, Locky and TeslaCrypt. This is popular between cybercriminals because it does not require user intervention, can reach potentially millions of users (through Google and other major advertising networks) and can cost as little as 30 cents per 1K impressions (Cox, 2015) so the return of investment is guaranteed. This malvertising strategy is relevant to price discrimination because criminals can bid for customers with a certain profile, belonging to a given demographic, or that have searched for certain keywords, thus basically abusing the ad network services as a proxy for cheap customer classification. No evidence of malvertising for price discrimination has been documented so far, but we will not be surprised if criminals start to employ it.

\subsection*{Bargaining}

The focus of our analysis so far has been on the criminals. But, is there any way for those targeted to counteract them? Trying to answer this question, we can look to bargaining theory. The classic model of bargaining assumes both parties, the criminal and the victim in our case, can make alternating offers (Rubinstein 1982). For
instance, the criminal sets a ransom, the victim then either pays the ransom or says how much she is willing to pay, then the criminal either accepts the counter-offer or sets a new ransom, and so on. It is known that this model is general enough to capture a wide range of realistic bargaining situations (Muthoo 1999). 

Suppose that $v_{i}$ is known by the criminals. Then the `predicted' outcome, the sub-game perfect Nash equilibrium (Muthoo 1999) is that the criminals will set an initial ransom of

\[
p_{i}=\left( \frac{1-\delta _{B}}{1-\delta _{A}\delta _{B}}\right)(v_{i}-c)+c
\]

\noindent where $\delta _{A}\leq 1$ is the discount factor of the criminals and $\delta _{B}\leq 1$ the discount factor of the victim. Informally, the discount factor measures the cost of delay in reaching an agreement. For instance, if the unit of time is one week, then the criminals would consider receiving \$$1$ in a weeks' time as the equivalent of receiving $\$\delta _{A}$ today. Similarly, the victim would consider receiving files worth $v_i$ in one weeks' time as equivalent to $\delta _{B} v_i$ today.

 \begin{figure}[!h]
 \begin{center}
 \includegraphics[scale=0.45]{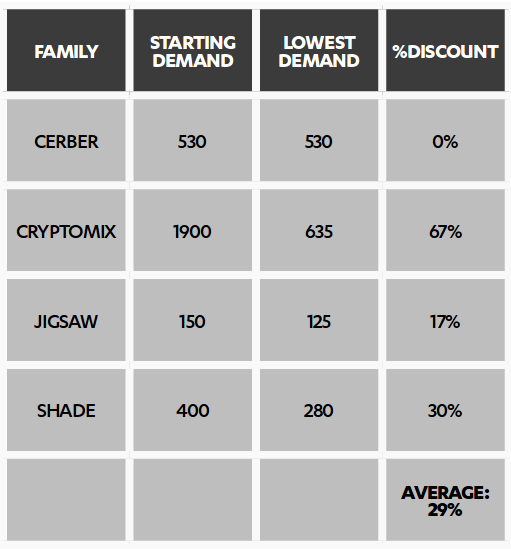} 
 \end{center}
 \caption{Some examples of ransomware open to bargaining, from (F-Secure, 2017)}
 \label{fig:bargaining}
 \end{figure}

The discount factors are assumed to be less than one because delay has an opportunity cost. The criminals, however, have relatively little to lose from a delay of, say, one week or one month in receiving the ransom given that their activities are likely to be ongoing over larger time periods. So, we would expect $\delta _{A} \approx 1$. The victim, however, will likely have more to lose from delay. For example, if they need immediate access to their files for work, or for filling in a tax return, etc. then a week or months delay may be very costly. We would expect, therefore, that $\delta _{B} < \delta _{A}$. Notice, though, that the value of $\delta _{B}$ is largely irrelevant if $\delta _{A} \approx 1$. If the discount factor of the criminals is close to one then, in equilibrium, the ransom will be close to the victim's willingness to pay. Moreover, the victim cannot gain from bargaining. 

In most real settings, the criminals have a strong bargaining position in the alternating offers game because they will generally be willing to wait. They can, though, strengthen their position even more by changing the rules of engagement. More specifically, the alternating offers game allows the victim to make a counter-offer. If, instead, the criminals make a take-it-or-leave-it offer to the victim then we have an ultimatum game (Binmore et al. 2002). The equilibrium prediction in the ultimatum game is that the ransom can be set arbitrarily close to $v_i$ (Muthoo 1999). In other words, the criminals can extract maximum profits. The ultimatum works by removing all bargaining power of the victim. It is important to clarify that extensive evidence suggests `unfair' offers in the ultimatum game are often rejected (e.g. Thaler 1988) and so the criminals may do well to set the ransom a little below $v_i$. The general point, however, remains that it is not in the interests of the criminals to encourage counter-offers. 

Let us now move on to the more realistic scenario where $v_{i}$ is not known by the criminal. The crucial concept here is the Coase conjecture (Coase 1972). To explain the issues, suppose that the criminals identify the optimal ransom $p$ (or its equivalent in the case of third degree price discrimination). Also suppose that they ask a person to pay the ransom $p$, and she does not pay. Given that she was not willing to pay a ransom of $p$, should the criminals come back with a lower ransom?

Intuition might suggest that they should lower the ransom because this increases their chances of making some profit, and if this was a one-off interaction that intuition would be correct\footnote{This would be an example of dynamic price discrimination. To give an analogous example consider a publisher who charges \$20 for a new released book and drops the price to \$5 after one year. The objective is to attract customers who were not willing to pay \$20 but might pay \$5.}. The criminals, however, clearly want to target more than one victim and this undermines the incentive to offer a `discount' to any one victim. Basically, if the criminals lower the ransom for one victim then every other victim is going to want the same discount\footnote{The nature of the good is important here. For instance, the value of a book will diminish quickly over time for some (because they want to be up to date with the latest release) and these will be willing to pay a high release price. For others, however, the value of a book does not diminish over time (because they just want to read a good book) and so they are willing to wait a year for a lower price. With ransomware it is likely that the value of files diminishes quickly for \textit{all} victims. This renders dynamic price discrimination unprofitable.}. The Coase conjecture formalizes this intuition and says that any anticipation of a fall in the ransom significantly diminishes the bargaining power of the criminals (Gul, Sonnenschein and Wilson 1986; Gul and Sonnenschein 1988).   

In order to maintain bargaining power the criminals need to credibly commit to not lowering the ransom over time (Kennan and Wilson 1993). In other words, the criminals need to set the ransom at the optimal level, and commit to no negotiation. The only way the criminals can credibly commit is through a reputation of being tough.

A particularly interesting real case following this approach is the Jigsaw ransomware, that after requesting a fixed ransom set at between $20-$150, depending on the versions, implemented an hourly definitive erase of a chunk of the victim's data, up to the 72 hours limit when all data would be permanently deleted (TrendMicro, 2016). Others, such as Razy 5.0 also claim to do so but do not really delete the data. In any case, it seems the threat is so credible that many (up to 33\%) of UK's larger business (250 or more employees) are holding bitcoins as part of their contingency plans, as revealed in a recent survey by Citrix (Parker, 2016). 

Provided there is a ‘never-ending’ stream of people being attacked, then it is possible to build a reputation for toughness. This is done by the criminals never giving-in, and for news of that to spread (possibly with help from the criminals themselves posting, say, ‘dummy’ messages on forums). Note that it is vital for the criminals bargaining position that there are always new victims on the horizon because this allows them to overcome the chain store paradox (Milgrom and Roberts 1982).The paradox, in this context, is that the threat to not lower the ransom over time would be incredible if there was a known end date to attacks. Basically, it would be in the interests of the criminals to lower prices for the last victim to be attacked, but then there is no point in not lowering prices for the penultimate victim, and the threat to not lower prices unravels.   

In Figure~\ref{fig:bargaining}, we can see examples of ransomware families that are not open to bargaining (Cerber, in the first row) and others that offer significant discounts. The simple phrase "That's too expensive, I don't really need the files that bad anyway" by an F-Secure employee led to these discounts (F-Secure, 2017).

According to our economic analysis, engaging into bargaining is, albeit tempting, not a good idea and thus the strategy followed by Cerber is better than that of the rest and will eventually generate their authors more profit and be adopted in future ransomware families.

\subsection*{Determinants of willingness to pay}

We have seen that the criminal's optimal strategy and profit is entirely dependent on people's willingness to pay to recover their files. We, therefore, finish this section by focusing on some variables that can influence this willingness to pay. We distinguish characteristics that are under the control of the criminals and those that are under the control of the victims. 

One thing the criminals control is whether to return files (or, alternatively, keys) to those who pay the ransom. The higher the probability of the files being recovered, the larger will be the willingness to pay, and the larger the ransom the criminals can set. In order to maximize profit the criminals should, therefore, always return files to people who pay the ransom, unless the cost of returning the files is very large. 

Another feature under the criminals' control is that of framing. For instance, the reflection effect means that people are, generally speaking, risk loving over losses and risk averse over gains (Pfleeger and Caputo 2012). This means that the willingness to pay will generally be larger if the criminals emphasize the sure loss of the files (`unless you pay the ransom your files are destroyed and lost forever') than if the criminals emphasize the possible gain of the files (`if you pay the ransom you recover your files'). In order to maximize profits, the criminals should frame the ransom demand with a focus on the sure loss of files. 

One thing under the control of victims is the `value' of the files. If a person has just backed up their files, then $v_{i}=0$. Clearly this is the easiest way to counter-act the criminals. The crucial and non-trivial point to emphasize, though, is that a person backing up her files benefits everyone who might be attacked, and not just herself. This is because the more people back up their files, the lower is $Q(p)$, the lower becomes the optimal ransom, and the less profit the criminals can make. In other words, a person's decision to back up their files generates a positive externality. 

Another thing under the control of people is whether or not to pay. The fewer people pay the ransom, the lower is $Q(p)$, the lower the optimal ransom, and the less profit the criminals can make. The decision to not pay also, therefore, generates a positive externality. The more a victim internalizes the two externalities we have mentioned, i.e. the more she takes into account the effect her actions will have on others, then the lower will be her willingness to pay. 

The preceding two paragraphs point to ways in which the impact of ransomware can be diminished. The idea that people need to back up their files is obvious. Less obvious, but potentially powerful, is the role of positive externalities. Strong social norms towards regularly backing up files and not paying a ransom (in the case the files were not backed up) greatly undermine the profit potential of cyber criminals.

\section*{Survey Results}

In order to obtain some preliminary estimates of people's willingness to pay to recover their files we conducted a face-to-face survey using standard contingent valuation techniques (Alberini and Kahn 2009). The survey was performed at the University of Kent (UK) during a celebration day on campus. A total of 149 respondents took part (54\% male, average age of $%
24$). Note that because we conducted the survey on a celebration day the respondents were primarily alumni or residents of the city (and not students). 

Half of those surveyed were asked the following two questions:\\

\noindent \textit{1. Suppose that because of a mistake you made, you have lost access to all of the files on your computer. The only way you can recover the files is to pay a private company who are experts in file recovery. What is the maximum amount you would pay to recover your files?
}\\
\noindent \textit{2. Suppose that your computer was infected by a virus which means you cannot access any of your files. The criminals responsible have been caught and you are now eligible for monetary compensation. How much money would you want to recompense you for the loss of files? Note that if your request is deemed too high the authorities will use a technique to recover your files and so 
you will receive your files but no compensation.
}\\
Question 1 directly elicits willingness to pay (WTP) - how much is someone willing to pay to recover their files. We know, however, that people have a tendency to understate willingness to pay. Question 2 addresses this problem by eliciting willingness to accept (WTA) - how much would someone need to be paid in order to compensate for the loss of files. In principle, WTP and WTA should be identical. Typically, however, one obtains a WTA-WTP disparity in that WTA is significantly higher than WTP (Horowitz and McConnell 2002). This difference can be anything from a factor of two to ten. The true valuation $v_{i}$ can reasonably be assumed to lie between an individuals' stated WTA and WTP. There are arguments to suggest that the true valuation will be, in fact, closer to WTA than to WTP (Bateman et al. 2005). \\
As a robustness check, the other half of those surveyed were asked the following questions:\\
\noindent \textit{3. Suppose that because of a mistake you made, you cannot access any of your
files. You have an insurance policy that means you are eligible for monetary
compensation. How much money would you want to recompense you for the loss
of files? Note that if your request is deemed too high, a technique will be
used to recover your files meaning you will receive your files but no
compensation.
}\\
\noindent \textit{4. Suppose that your computer was infected by a virus which means you cannot
access any of your files. The only way you can recover your files is if you pay
a fee to the criminals. If you can be certain that your files will be
returned, what is the maximum you would pay to recover your files?
}\\
Note that question 3 elicits WTA while question 4 elicits WTP. Reversing the
order of questions allows us to check that respondents' stated WTA and WTP is
not influenced by the order in which the questions are asked. Moreover, the
framing varies between questions 1 and 4, and 2 and 3 in terms of who was
responsible for the loss of files and who will be paid or recompensed.
Ideally we would want answers to questions 1 and 4 to be similar, and those
to questions 2 and 3 to be similar. This was indeed the case and so, in the following, we shall
solely report stated WTP (questions 1 and 4) and WTA (questions 2 and 3).\footnote{We performed the non-parametric Wilcoxon rank-sum test that answers to questions 1 and 4 are from the same distribution ($p = 0.55$) and that the answers to questions 2 and 3 are from the same distribution ($p = 0.22$).}

\begin{figure}[!h]
 \begin{center}
 \includegraphics[scale=1.2]{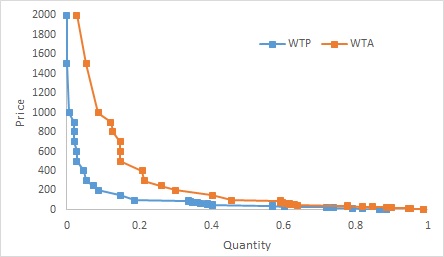} 
 \end{center}
 \caption{Demand curve elicited using WTA and WTP}
 \label{fig:demandcurve}
 \end{figure}

In economics, it is conventional to plot the inverse demand curve $%
Q^{-1}\left( p\right)$. Figure~\ref{fig:demandcurve} plots the inverse demand curve using the elicited
values of WTA and WTP with total demand normalized to 1. Note that price is in sterling and at the time of the survey the exchange was around \$1.40 per \pounds1.00. As expected, the WTA (mean \pounds 547) exceeds the WTP (mean \pounds 276) resulting in higher demand when using the WTA as compared to the WTP. For instance, 20\%
of those surveyed were willing to accept at least \pounds 400 and pay up to 
\pounds 100. This is a WTA-WTP disparity of factor four. 

%\bigskip \includegraphics[width = \textwidth]{demandfn.jpg}
%\bigskip \noindent Figure 1: Demand curve elicited using WTA and WTP.

\bigskip As discussed above, elicited WTA can be seen as a better measure of true
valuation. In order to calculate the profit maximizing ransom we fitted a (six degree) polynomial to the raw demand function elicited using WTA\footnote{%
The fitted equation is $%
p=-37950Q^{5}+116699Q^{4}-137561Q^{3}+77678Q^{2}-21367Q+2472.1$.}. From this fitted demand function, we could then calculate marginal revenue (MR) using equation%
\[
MR(Q)=\frac{dp(Q)}{dQ}Q+p(Q).
\]
The optimal ransom demand is found where marginal revenue equals marginal
cost. It seems reasonable to assume that the marginal cost to the criminals of
dealing with an additional victim is near zero. The optimal ransom is,
therefore, found by setting $MR(Q)=0$.

\begin{figure}[!h]
 \begin{center}
 \includegraphics[scale=1.2]{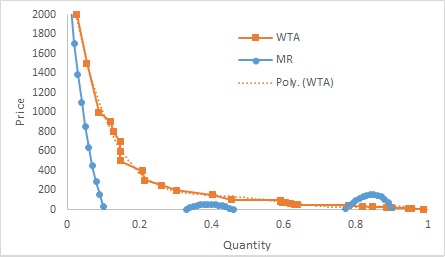} 
 \end{center}
 \caption{Demand curve elicited using WTA, and marginal revenue (MR).}
 \label{fig:demandcurve2}
 \end{figure}

Figure~\ref{fig:demandcurve2} plots the raw demand function (the same as in Figure 1), the fitted demand function and marginal
revenue. Note that there are 5 values of $Q$ for which $MR(Q)=0$. Only one of these is the global optimum and it is simple to show that this is the smallest $Q$ where $MR(Q)=0$. This gives an optimal ransom of
around \pounds 950. It is predicted that, in this setting, around 10\% of victims will pay. An
interpretation of this finding is that it is in the criminal's interest to
target high value victims that are willing to pay \pounds 1000 or above. For
instance, at the optimum value the expected profit of the criminals is \pounds 99
per victim (because a little over 10\% of victims will pay \pounds 950). If the
ransom is dropped to, say, \pounds 150 then over 40\% of victims will pay
but this only results in a profit per victim of around \pounds 60. 

%\bigskip \includegraphics[width = \textwidth]{MR.jpg}
%\bigskip \bigskip \noindent Figure 2: Demand curve elicited using WTA, and marginal revenue (MR).

\bigskip Clearly, our survey results can be improved, for example with a larger and more representative sample, etc. So our findings are not definitive, and more evidence to support them is needed. 

Our results do, however, demonstrate the potential for a method to calculate an optimal ransom. Moreover, we would not be surprised to find that the optimal ransom is much higher than what is currently seen in most ransomware. As our results illustrate, if there are a few people willing to pay a high ransom, then it may well be optimal to set a much higher ransom and accept that many victims will not pay up\footnote{It may be the case that cybercriminals are discovering this fact, and ransom prices will continue to increase until reaching our estimated optimal level. \textit{The average ransom demand has more than doubled from $294 at the end of 2015 to $679} writes Hyacinth Mascarenhas
on July 21, 2016, for International Business Times, following a study by Symantec.}. 

If it is the case that the optimal (uniform) price would result in many victims being unwilling to pay then it is especially in the criminal's interest to price discriminate. Ideally the criminals would want to identify those willing to pay a high ransom and ask a high ransom while making a lower ransom demand to others. This, though, requires being able to discern those willing to pay a high ransom. We find in our data that the willingness to pay is on average higher for women (mean \pounds 326) than men (mean \pounds 233) and increasing with age (correlation coefficient 0.07). These differences are, however, not statistically significant and so nothing more than suggestive. Again, however, they demonstrate the potential to estimate to what extent willingness to pay depends on discernible characteristics of an individual. This is key if the criminals are to effectively utilize price discrimination.    

\section*{Conclusions}

There can be no doubt that ransomware is already a serious security threat to individuals, firms and beyond. It is natural, though, to expect that this threat will evolve over time as the criminals learn from experience and refine their techniques. Indeed, with something in place similar to a selection of the fittest, it is very likely that ransomware will slowly evolve towards the optimal strategies that can be known in advance thanks to Economics and Game Theory. 

While data on current activity is somewhat limited, we would suggest that the techniques currently being used by the criminals are relatively unsophisticated. There certainly seems to exist ample scope for them to refine their techniques, notably for determining the optimal ransom and to make use of price discrimination. Crucially, refining strategies in the ways we have suggested in this work could lead to dramatic increases in profits at relatively little costs. In particular, the criminals should have at their disposal a wealth of data regarding the willingness of victims to pay a ransom. With only a rudimentary analysis of this data, they could almost certainly obtain higher profits. A more thorough analysis could push  profits up even more. With strong incentives for the criminals to innovate, they are surely going to do so.

In this work we focused exclusively on the criminals economic strategy. There is also scope for innovation in terms of the behavioral strategy employed (Pfleeger and Caputo 2012). To reiterate the earlier example, we know that individuals are averse to losses (Kahneman et al. 1991). This would suggest that a victim would be willing to pay a higher ransom in order to avoid the loss of their files than they would to gain access to files. Moreover, they would likely pay more if they are primed to think about the files they will potentially lose. The key point to appreciate here is that the optimal strategy needs to take account of both economic and behavioral considerations. It is in the criminal's interests to use both a behavioral strategy that maximizes willingness to pay and then to use an economic strategy that maximizes profits (given willingness to pay).

Is it all bad news if the criminals do indeed develop better techniques for maximizing profit in the future? Economic theory tells us that a more efficient strategy on their part can, again counter intuitively, increase overall welfare. The basic reason for this in the case of ransomware is that many victims may end up losing their files if criminals employ an unsophisticated strategy, while most will recover their files if they employ a better strategy. After all, it is in the interest of the criminals that victims pay up, and victims are only going to do that if they have a chance of getting their files back and the ransom price is right. 

It does not, therefore, immediately follow that the average victim will suffer if the criminals develop a better, sounder economic strategy. On the other hand, it is very important to highlight the positive externalities globally created by victims backing-up their data and refusing to pay ransoms. Possibly, a stronger case for both of these behaviors should be made in an effort to effect a change in victim's actions. 
%Clearly, however, there are negative consequence in criminals extracting a higher profit. 
Finally, the analysis in our work suggests that profits from ransomware will likely increase in the future, as criminals will slowly discover and converge towards optimal strategies, and increase ransom prices closer to their optimal value. 

%Comment on the effect of contract-based cryptocurrencies such as Ethereum, as mentioned by Matthew Green at https://blog.cryptographyengineering.com/2017/02/28/the-future-of-ransomware/

%Comment briefly on this piece, on malware mining monero, as a way to show cybercriminals are taken an increased interest in this alternative cryptocurrency: https://news.bitcoin.com/monero-mining-malware/

%Add my table with average ransoms and other characteristics of a bunch of ransomware. It's in my dropbox, complete it a bit.

\section*{References}

\smallskip \noindent Alberini, A. \& Kahn, J. (2009). \textit{Handbook on contingent valuation}. Edward Elgar. 

\smallskip \noindent Anderson, R., Barton, C., Böhme, R., Clayton, R., Van Eeten, M., Levi, M., Moore, T., \& Savage, S. (2013). Measuring the cost of cybercrime. \textit{The Economics of Information Security and Privacy}. Springer Berlin Heidelberg.

\smallskip \noindent Bateman, I., Kahneman, D., Munro, A., Starmer, C., \& Sugden, R. (2005). Testing competing models of loss aversion: an adversarial collaboration. \textit{Journal of Public Economics}, 89(8), 1561-1580.

\smallskip \noindent Binmore, K., McCarthy, J., Ponti, G., Samuelson, L., \& Shaked, A. (2002). A backward induction experiment. \textit{Journal of Economic theory}, 104(1), 48-88.

\smallskip \noindent Coase, R. H. (1972). Durability and monopoly. \textit{Journal of Law \& Economics}, 15: 143-149.

\smallskip \noindent Gul, F., \& Sonnenschein, H. (1988). On delay in bargaining with one-sided uncertainty. \textit{Econometrica}, 56: 601-611.

\smallskip \noindent Gul, F., Sonnenschein, H., \& Wilson, R. (1986). Foundations of dynamic monopoly and the Coase conjecture. \textit{Journal of Economic Theory}, 39: 155-190.

\smallskip \noindent Horowitz, J. K., \& McConnell, K. E. (2002). A review of WTA/WTP studies. \textit{Journal of Environmental Economics and Management}, 44(3), 426-447.

\smallskip \noindent Kennan, J., \& Wilson, R. B. (1993). Bargaining with private information. \textit{Journal of Economic Literature}, 31: 45-45.

\smallskip \noindent Kahneman, D., Knetsch, J. L., \& Thaler, R. H. (1991). Anomalies: The endowment effect, loss aversion, and status quo bias. \textit{The Journal of Economic Perspectives}, 5(1), 193-206.

\smallskip \noindent Milgrom, P., \& Roberts, J. (1982). Predation, reputation, and entry deterrence. \textit{Journal of Economic Theory}, 27(2), 280-312.

\smallskip \noindent Muthoo, A. (1999). \textit{Bargaining theory with applications}. Cambridge University Press.

\smallskip \noindent Pepall, L., Richards, D. J., \& Norman, G. (2008). \textit{Industrial organization: Contemporary theory and empirical applications}. Blackwell: Oxford.

\smallskip \noindent Pfleeger, S. L., \& Caputo, D. D. (2012). Leveraging behavioral science to mitigate cyber security risk. \textit{Computers \& security}, 31(4), 597-611.

\smallskip \noindent Rubinstein, A. (1982). Perfect equilibrium in a bargaining model. \textit{Econometrica}, 50: 97-109.

\smallskip \noindent Selten, R. (1977). A simple game model of kidnapping. \textit{Lecture Notes in Economics and Mathematical Systems} 141: pp 139-155.

\smallskip \noindent Thaler, R. H. (1988). Anomalies: The ultimatum game. The Journal of Economic Perspectives, 2(4), 195-206.

\smallskip \noindent Varian, H. R. (1989). Price discrimination. \textit{Handbook of industrial organization Volume 1}, R. Schmalensee and R. Willig (Eds.) Elsevier: North Holland, pp. 597-654.

\smallskip \noindent Young, A., \& Yung, M. (1996). Cryptovirology: Extortion-based security threats and countermeasures. \textit{Security and Privacy Proceedings IEEE Symposium}. IEEE.

\smallskip \noindent Emm D. (2008). Cracking the code: The history of Gpcode. \textit{Computer Fraud \& Security
Volume 2008, Issue 9, September 2008, pages 15-–17}. Elsevier.

\smallskip \noindent Jouhal J. (2017). The rise of ransomware and how to avoid being held hostage. \textit{New Statesman, Spotlight on cybersecurity, pages 8--9, February 2017}.

\smallskip \noindent  Sinitsyn, F. (2015). A flawed ransomware encryptor \textit{Securelist Blogpost, April 8, 2015}.\\ https://securelist.com/blog/research/69481/a-flawed-ransomware-encryptor/

\smallskip \noindent  Hahn, K. (2016). The Rise of Low Quality Ransomware \textit{G-DATA Security blog, 21st September 2016}.\\ https://blog.gdatasoftware.com/2016/09/29157-the-rise-of-low-quality-ransomware

\smallskip \noindent Atanasova, S. (2016). The Shade Ransomware with New RAT Features to Determine Worthwhile Victims \textit{Virus Guides' Computer Security News of August 12, 2016}.\\ 
http://virusguides.com/shade-ransomware-new-rat-features-determine-worthwhile-victims/

\smallskip \noindent Abrams, L. (2016). Fantom Ransomware derives Ransom Amount and Address from Filename \textit{Bleeping Computer News of September 21, 2016}.\\ 
https://www.bleepingcomputer.com/news/security/fantom-ransomware-derives-ransom-amount-and-address-from-filename/
ransomware-hit-you

\smallskip \noindent  TrendMicro (2016). Ransom Notes: Know What Ransomware Hit You \textit{TrendMicro Security News of July 19, 2016}. https://www.trendmicro.com/vinfo/us/security/news/cybercrime-and-digital-threats/ransom-notes-know-what-

\smallskip \noindent F-Secure (2017)
\textit{State of Cyber Security in 2017}.\\ 
https://www.f-secure.com/documents/996508/1030743/cyber-security-report-2017

\smallskip \noindent Cox, J. (2015)
\textit{Malvertising, the Hack that Infects Computers Without a Click}.
https://www.wired.com/2015/12/hacker-lexicon-malvertising-the-hack-that-infects-computers-without-a-click/
\\ 

\smallskip \noindent  Parker, L. (2016). Large UK businesses are holding bitcoin to pay ransoms.\textit{ Bravenewcoin.com News, 9 June 2016}.
http://bravenewcoin.com/news/large-uk-businesses-holding-bitcoin-to-pay-ransoms/
\\

\subsubsection*{Acknowledgements\\}

\begin{center}
\begin{tabular}{ c c }
  
   \begin{minipage}{.15\textwidth}
      \includegraphics[width=\linewidth, height=12mm]{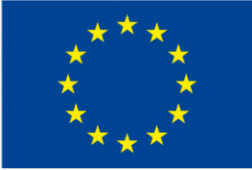}
    \end{minipage}
  
  &   

\begin{minipage}{.6\textwidth}
  
This project has received funding from the European Union’s Horizon 2020 research and innovation programme, under grant agreement No.700326 (RAMSES project). The authors also want to thank EPSRC for project EP/P011772/1, on the EconoMical, PsycHologicAl and Societal Impact of RanSomware (EMPHASIS), which supported this work.
  \end{minipage}

\end{tabular}
\end{center}

\end{document}